\documentclass[aps,prl,10pt,twocolumn,superscriptaddress,balancelastpage,showpacs,reprint,notitlepage]{revtex4-2}
\usepackage[english]{babel}
\usepackage[utf8]{inputenc}
\usepackage[colorinlistoftodos, color=green!40, prependcaption]{todonotes}

\newcommand{\qq}{{\bm q}}
\newcommand{\pp}{{\bm p}}
\newcommand{\rr}{{\bm r}}

\newcommand{\UU}{{\bm U}}
\newcommand{\aapp}{\alpha_{\mathrm{app}}}

\usepackage{centernot}
\usepackage{graphicx}
\usepackage{bm}
\usepackage{amsmath}
\usepackage{times}
\usepackage{amssymb}
\usepackage{mathrsfs}
\usepackage{color}
\usepackage{url}
\usepackage{version}
\usepackage{siunitx}
\usepackage[pdftex,colorlinks=true,pdfstartview=FitV,linkcolor= linkcolor,citecolor= linkcolor,urlcolor= linkcolor,hyperindex=true,hyperfigures=false]{hyperref}
\usepackage{balance}
\usepackage{lastpage}
\usepackage{parskip}
\usepackage{siunitx}
\usepackage[paper=a4paper,margin=1in]{geometry}
\usepackage[caption=false]{subfig}

\definecolor{linkcolor}{rgb}{0,0,0.6}

\DeclareSIUnit{\angstrom}{\textup{\AA}}

\begin{document}

\title{Pseudo-giant number fluctuations and nematic order in microswimmer suspensions}

\author{Ismail El Korde}
\affiliation{Division of Physical Chemistry, Lund University, P.O. Box 124, S-221 00 Lund, Sweden}

\author{D\'ora B\'ardfalvy}
\affiliation{Division of Physical Chemistry, Lund University, P.O. Box 124, S-221 00 Lund, Sweden}

\author{Jason M. Lewis}
\affiliation{Division of Physical Chemistry, Lund University, P.O. Box 124, S-221 00 Lund, Sweden}

\author{Alexander Morozov}
\affiliation{School of Physics and Astronomy, The University of Edinburgh, James Clerk Maxwell Building, Peter Guthrie Tait Road, Edinburgh, EH9 3FD, United Kingdom}

\author{Cesare Nardini}
\affiliation{Service de Physique de l'\'Etat Condens\'e, CNRS UMR 3680, CEA-Saclay, 91191 Gif-sur-Yvette, France}
\affiliation{Sorbonne Universit\'e, CNRS, Laboratoire de Physique Th\'eorique de la Mati\`ere Condens\'ee, LPTMC, F-75005 Paris, France.}

\author{Joakim Stenhammar}
\email{joakim.stenhammar@fkem1.lu.se}
\affiliation{Division of Physical Chemistry, Lund University, P.O. Box 124, S-221 00 Lund, Sweden}

\date{\today} 

\begin{abstract}
\noindent Giant number fluctuations (GNFs), whereby the standard deviation $\Delta N$ in the local number of particles $\langle N \rangle$ grows faster than $\sqrt{\langle N \rangle}$, are a hallmark property of dry active matter systems with orientational order, such as a collection of granular particles on a vibrated plate. This contrasts with momentum-conserving (``wet'') active matter systems, such as suspensions of swimming bacteria, where no theoretical prediction of GNFs exist, although numerous experimental observations of such enhanced fluctuations have been reported. In this Letter, we numerically confirm the emergence of super-Gaussian number fluctuations in a 3-dimensional suspension of pusher microswimmers undergoing a transition to collective motion. These fluctuations emerge sharply above the transition, but only for sufficiently large values of the bacterial persistence length $\ell_p = v_s / \lambda$, where $v_s$ is the bacterial swimming speed and $\lambda$ the tumbling rate. Crucially, these ``pseudo-GNFs'' differ from true GNFs, as they only occur on length scales shorter than the typical size $\xi$ of nematic patches in the collective motion state, which is in turn proportional to the single-swimmer persistence length $\ell_p$. Our results thus suggest that observations of enhanced density fluctuations in biological active matter systems actually represent transient effects that decay away beyond mesoscopic length scales, and raises the question to what extent ``true'' GNFs with universal properties can exist in the presence of fluid flows. 
\end{abstract}

\maketitle

\noindent Collective motion in active matter typically arises due to the coupling between activity and orientational order~\citep{Marchetti:RMP:2013}. One generic feature of orientationally ordered flocks is the emergence of anomalously large fluctuations in the particle density, so-called giant number fluctuations (GNFs), whereby the standard deviation $\Delta N$ in the local number of particles $N$ grows as $ \langle N \rangle^{\alpha}$ with $\alpha > 1/2$~\citep{Sriram:AnnuRev:2010}. Experimentally, the emergence of GNFs in dry active matter was first demonstrated in a collection of rods on a vibrating plate~\citep{Narayan:Science:2007}, where collective motion arises from the coupling between activity-driven particle currents and (quasi-)long range order caused by particle collisions, and has since then been experimentally~\citep{Sood:NatCommun:2014,Chate:PRL:2010} and numerically~\citep{Chate:PRL:2006,Chate:PRE:2008} verified in both nematic and polar flocks. The emergence of GNFs in dry flocks can be understood by calculating the autocorrelation of density fluctuations in the orientationally ordered state~\citep{Sriram:EPL:2003,Toner:JCP:2019}. In the nematic case, the ensuing analysis shows that, at linear level, the variance $\langle \delta n(\qq,t)\delta n(-\qq,t)\rangle$ in the local number density $n$ of the ordered flock diverges as $q^{-2}$ for small $q = |\qq|$, which, in 2 dimensions, yields a linear growth of $\Delta N$ with $\langle N \rangle$, \emph{i.e.}, $\alpha = 1$. This picture is changed when accounting for the effect of nonlinearities~\citep{Marchetti:PRE:2018}, which lead to non-universal exponents that depend on the microscopic model parameters. 

In wet active matter, the situation is more complex: Since the nematically ordered state is generically unstable for finite wavenumbers in the presence of fluid flows~\citep{Simha:PRL:2002}, the route used in the dry case to calculate $\langle \delta n(\qq,t)\delta n(-\qq,t)\rangle$, based on a computation at linear level followed by the treatment of non-linearities within a renormalization group framework~\citep{Sriram:EPL:2003,Marchetti:PRE:2018}, is no longer accessible. Therefore, no quantitative predictions exist for the emergence of GNFs in bulk wet active nematics, although they have been predicted for nematic flocks moving through a fluid in contact with a substrate~\citep{Maitra:PNAS:2018}. There are however numerous experimental~\citep{Cheng:SoftMatter:2021,Silberzan:SoftMatter:2014,Sano:PRE:2017,Peruani:PRL:2012,Swinney:PNAS:2010} (and some numerical~\citep{Saintillan:Interface:2012}) observations of enhanced concentration fluctuations in various biological incarnations of wet active matter, suggesting that such fluctuations are indeed a common feature also in the presence of momentum conservation; it is however difficult to assess whether these observations represent ``true'' GNFs with asymptotically valid power laws and associated universal exponents or merely transient effects where fluctuations appear non-Gaussian because distributional convergence has not yet been reached. 

The archetypical example of collective motion in wet active matter is the transition to so-called bacterial turbulence in dilute suspensions of rear-actuated ``pusher'' microorganisms such as \emph{E. coli} and \emph{B. subtilis}~\citep{Clement:NJP:2014,Aranson:PRL:2012,Goldstein:PRL:2013,Wensink:PNAS:2012}. In a 3-dimensional, unbounded suspension of pusher microswimmers at density $n$, this transition is predicted to occur whenever ~\citep{Hohenegger:PRE:2010,Subramanian:JFM:2009,Stenhammar:PRL:2017}
\begin{equation}\label{eq:n_c_inf}
n > n_c^{\infty} = \frac{5\lambda}{\kappa},
\end{equation}
where $\kappa > 0$ is the dipole strength and $\lambda$ is the rate of bacterial reorientation (``tumbles''). The critical density in Eq.~\eqref{eq:n_c_inf} represents the long-wavelength destabilisation of the homogeneous and isotropic suspension caused by bacterial reorientation due to flow alignment. Since the dipolar flow field is fore-aft symmetric, the ensuing orientationally ordered state exhibits nematic symmetry. This state however in turn destabilises due to the presence of fluid flows~\citep{Simha:PRL:2002}, resulting in a highly nonlinear, turbulent state characterised by short-ranged nematic order and long-range chaotic flows~\citep{Bardfalvy:SoftMatter:2019,Saintillan:PoF:2008}. The fore-aft symmetry is broken by the presence of self-propulsion, whereby each swimmer moves along its director $\pp$ with constant speed $v_s$. Crucially, however, $v_s$ does \textit{not} affect the mean-field critical density in Eq.~\eqref{eq:n_c_inf}~\citep{Skultety:PRX:2020}; this observation is far from trivial, as exemplified by the emergence of collective motion near a confining boundary, where $v_s$ enters the expression for $n_c$ even at mean-field level~\citep{Skultety:JFM:2024,Bardfalvy:CommPhys:2024}.

In this Letter, we numerically confirm the experimental observation that super-Gaussian number fluctuations occur generically above the transition to collective motion in microswimmer suspensions. Unlike in dry active systems, however, these only occur on length scales small compared to the bacterial persistence length $\ell_p = v_s / \lambda$, beyond which number fluctuations start decaying towards Gaussian behaviour. By analysing the range of orientational order in the collective motion regime, we show that the linear size $\xi$ of nematically ordered patches is also proportional (and comparable) to $\ell_p$, establishing a direct link between nematic order and non-Gaussian number fluctuations. Our results demonstrate that true GNFs, in the sense of being a large-scale property exhibiting universal exponents, are generically suppressed in the presence of fluid flow, providing a new perspective on experimental observations of anomalous number fluctuations in wet active systems~\citep{Cheng:SoftMatter:2021,Silberzan:SoftMatter:2014,Sano:PRE:2017,Peruani:PRL:2012,Swinney:PNAS:2010}. 

\begin{figure}[h]
  \begin{center}
  \includegraphics[width=65mm, clip, trim=0.4cm 0.6cm 0.4cm 0.4cm]{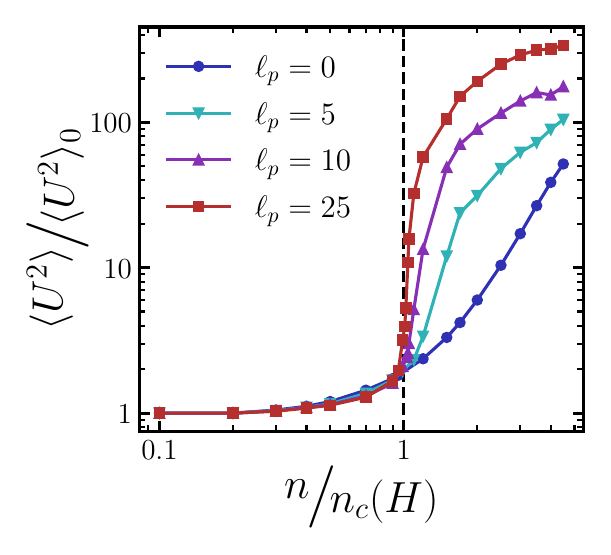}
\caption{Fluid velocity variance $\langle U^2 \rangle$ normalised by its corresponding value in a non-interacting suspension, $\langle U^2 \rangle_0$, for various persistence lengths $\ell_p$. The microswimmer density $n$ is normalised by the critical density $n_c(H)$ from Eqs.~\eqref{eq:n_c_inf} and~\eqref{eq:n_c_H}. Note that increasing the swimming speed, measured through $\ell_p$, sharpens the transition by suppressing pretransitional swimmer-swimmer correlations.}\label{fig:U2}
  \end{center}
\end{figure}

We consider a dilute, 3-dimensional suspension of pusher microswimmers at nominal number density $n$ moving through an incompressible, viscous fluid with viscosity $\mu$ and mass density $\rho$. Each swimmer is modeled as an extended force dipole of magnitude $\kappa = F\ell / \mu$ acting on the fluid, with $F$ the force magnitude and $\ell$ the dipole length. The position $\rr_i$ and orientation $\pp_i$ of swimmer $i$ evolve according to
\begin{align}
\label{eq:rdot} \dot{\rr}_i &= v_s \pp_i + \UU(\rr_i), \\
\label{eq:pdot} \dot{\pp}_i &= (\mathbb{I}-\pp_i\pp_i) \cdot \nabla \UU(\rr_i) \cdot\pp_i,
\end{align}
where $\UU(\rr_i)$ is fluid velocity measured at the position of the swimmer body, $v_s$ is the (constant) swimming speed, and $\mathbb{I}$ is the unit tensor. In addition to the rotation by the ambient fluid described by Eq.~\eqref{eq:pdot}, each swimmer randomises its swimming direction (``tumbles'') with average frequency $\lambda$. The temporal evolution of the flow field $\UU(\rr,t)$ was solved using a point-force implementation of the lattice Boltzmann (LB) algorithm described previously~\citep{Nash:PRE:2008,Nash:PRL:2010,Bardfalvy:SoftMatter:2019,Stenhammar:PRL:2017}, using a cubic box of linear size $H$ with periodic boundaries; parameter values and technical details about the simulations are provided as End Matter. 

We begin by characterising the transition from disordered swimming to collective motion, and in particular the effect of varying swimming speed $v_s$. As an order parameter, we choose the reduced space- and time-averaged fluid velocity variance, $\langle U^2 \rangle / \langle U^2 \rangle_0$, where $\langle U^2 \rangle_0$ is the corresponding value measured in suspension of non-interacting microswimmers. In Fig.~\ref{fig:U2}, we plot this quantity as a function of the reduced microswimmer density $n / n_c(H)$, where $n_c(H)$ is the analytically known critical density in a finite system of linear size $H$~\citep{Poon:PNAS:2020}:
\begin{equation}\label{eq:n_c_H}
n_c (H) = n_c^{\infty} \left[ 1 + \frac{3\pi}{5}\frac{\ell_p}{H} + \frac{4\pi^2}{5}\left( \frac{\ell_p}{H}\right)^2 \right].
\end{equation}
Eq.~\eqref{eq:n_c_H} encodes the fact that the mean-field hydrodynamic instability leading to bacterial turbulence occurs at the longest length-scale available, which, in a system of linear size $H$, corresponds to a critical wave vector $k_c = 2\pi / H$. Using this correction, the results in Fig.~\ref{fig:U2} confirm the theoretical prediction from~\citep{Skultety:PRX:2020} that faster swimming effectively sharpens the transition to collective motion by suppressing pretransitional swimmer-swimmer correlations  without shifting the critical density. This sharpening is therefore \textit{not} a finite-size effect: As we show in~\citep{SI} (see also~\citep{Stenhammar:PRL:2017,Bardfalvy:SoftMatter:2019}), increasing $H$ does not lead to a sharpening of the transition, unlike what one would expect from a critical phase transition. The analysis in Fig.~\ref{fig:U2} also shows that the theoretically predicted critical density given by Eqs.~\eqref{eq:n_c_inf} and~\eqref{eq:n_c_H} coincides with the sharp increase in $\langle U^2 \rangle$. Even though the transition to collective motion has been numerically studied numerous times using  continuum~\citep{Saintillan:PoF:2008} and particle-resolved~\citep{Saintillan:Interface:2012,Bardfalvy:SoftMatter:2019} simulations, the results in Fig.~\ref{fig:U2} constitute the first quantitative verification of the prediction in Eq.~\eqref{eq:n_c_inf}. As seen in the snapshots in End Matter, changing $\ell_p$ leads to visible differences in the collective steady states: For $\ell_p = 0$, the flow field appears random with very short-ranged correlations even for $n / n_c = 2$, while faster swimming, corresponding to larger $\ell_p$, leads to increasingly large-scale flow structures in the velocity field. 

\begin{figure*}[hbt!]
\centering
 \subfloat{\includegraphics[width=75mm, clip, trim=0.25cm 0.25cm 0.22cm 0.22cm]{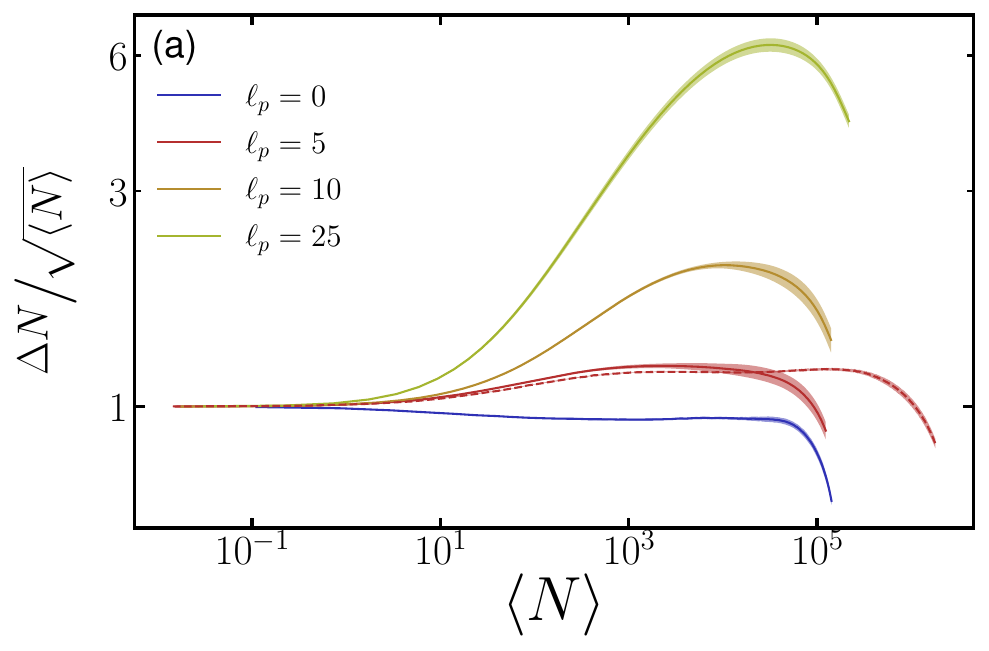}}\quad
 \subfloat{\includegraphics[width=75mm, clip, trim=0.5cm 0.5cm 0.4cm 0.4cm]{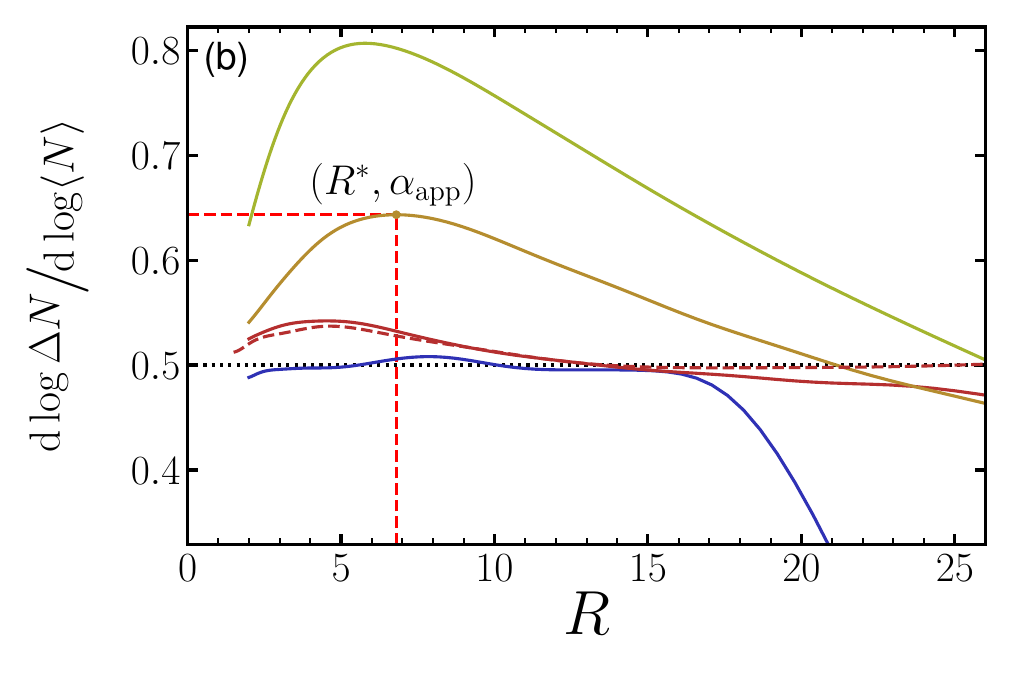}} \\
 \subfloat{\includegraphics[width=75mm, clip, trim=0.5cm 0.5cm 0.4cm 0.5cm]{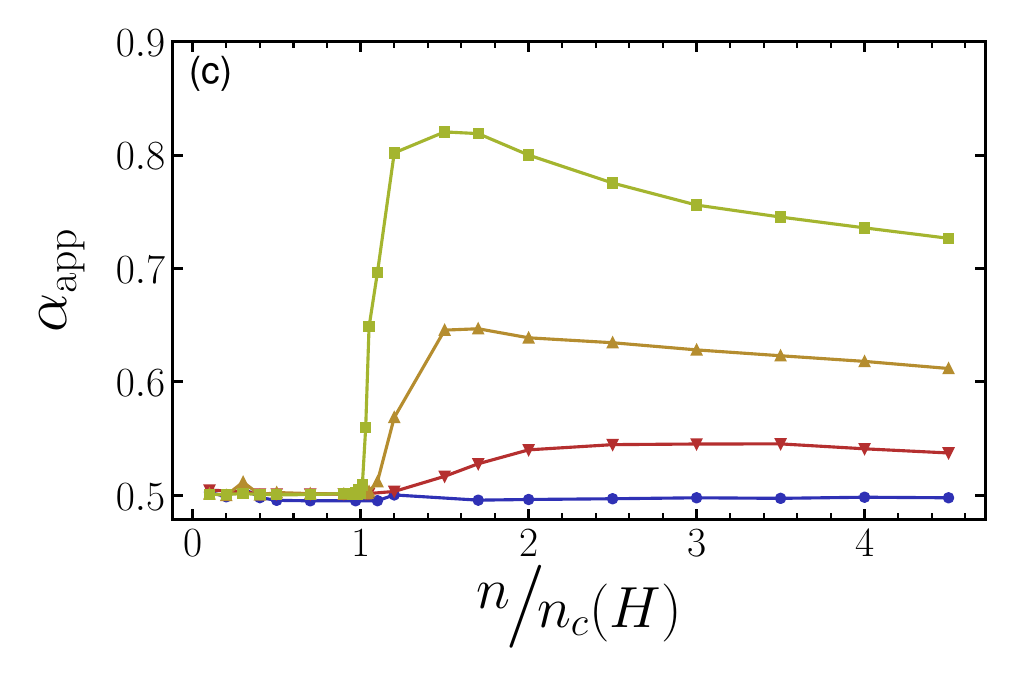}} \quad
 \subfloat{\includegraphics[width=75mm, clip, trim=0.5cm 0.5cm 0.4cm 0.5cm]{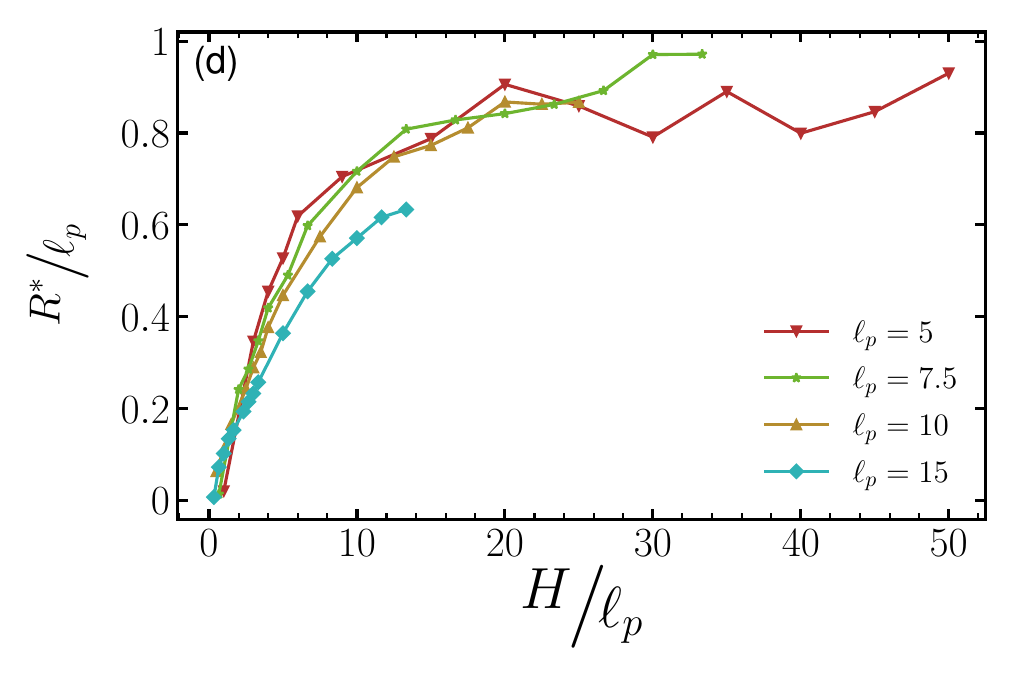}}
 \caption{
 (a) Reduced standard deviation $\Delta N / \sqrt{\langle N \rangle}$ of the average number of particles $N$ in a spherical volume as a function of $\langle \Delta N \rangle$ obtained for $n / n_c = 2.0$ and various $\ell_p$, as indicated. The dashed line for $\ell_p = 5$ shows the same quantity for a larger system size, $H=250$, showing the development of an extended plateau of Gaussian fluctuations for intermediate length scales. The shaded area denotes one standard deviation, estimated by averaging each run over ten sub-batches. 
 (b) The derivative $\mathrm{d} \log \Delta N /  \mathrm{d} \log \langle N \rangle$ plotted as a function of the radius $R$ of the sampling volume. The maximum in $\mathrm{d} \log \Delta N /  \mathrm{d} \log \langle N \rangle$ defines the apparent GNF exponent $\aapp$, and its location defines $R^*$. 
 (c) $\aapp$ as a function of the reduced microswimmer density $n / n_c$. Notice that non-Gaussian fluctuations emerge at the transition to collective motion only for sufficiently large $\ell_p$ and are completely absent for shakers.  
 (d) $R^* / \ell_p$ as a function of the reduced system size $H / \ell_p$ for $n / n_c = 2.0$. The data collapse shows that, for sufficiently large system sizes $H \gg \ell_p$, $R^*$ is controlled by $\ell_p$; hence, the apparently non-Gaussian fluctuations are only transient.}\label{fig:GNFs} 
\end{figure*}

To quantify the emergence of non-Gaussian number fluctuations at the transition to collective motion, we first consider the spatiotemporal average $\langle N \rangle$ and the corresponding standard deviation $\Delta N$ of the number of particles in randomly sampled spherical volumes of varying radius $R$. For $n > n_c$ and $\ell_p > 0$, it is clear from the results in Fig.~\ref{fig:GNFs}a that $\Delta N$ grows faster than $\sqrt{\langle N \rangle}$, indicating the emergence of GNFs in the collective motion state. For shakers ($\ell_p = v_s = 0$), no such enhanced fluctuations are seen, which, as was first noted in~\citep{Saintillan2015}, is a consequence of the fact that the local polar order which drives density fluctuations is strictly zero for shakers, since there is no term in the microscopic equations of motion that break fore-aft symmetry. For finite $\ell_p$, $\Delta N / \sqrt{\langle N \rangle}$ has a non-monotonic behaviour, with significantly enhanced number fluctuations for small and intermediate $\langle N \rangle$, followed by a fast decrease for larger $\langle N \rangle$. This behaviour is qualitatively similar to previous experimental~\citep{Cheng:SoftMatter:2021,Sano:PRE:2017,Narayan:Science:2007,Swinney:PNAS:2010,Peruani:PRL:2012,Sood:NatCommun:2014,Chate:PRL:2010} and numerical~\citep{Chate:PRL:2006,Chate:PRE:2008} observations of GNFs, and can partially be explained by the unavoidable damping of number fluctuations at large length scales due to global particle conservation; as we will see in the following, this is however not the primary mechanism at work here. That the underlying physics is more complex than just mass conservation is already visible from the two curves for $\ell_p = 5$ in Fig.~\ref{fig:GNFs}a and b: These show that increasing the system size from $H = 100$ to $H = 250$ does \textit{not} extend the region of enhanced number fluctuations as expected for true GNFs, but rather leads to the development of an intermediate plateau with Gaussian fluctuations preceding the decay due to finite $H$. 

Due to the difficulty in consistently fitting a power-law to the very limited data range, in Fig.~\ref{fig:GNFs}b we instead approximate the derivative $\mathrm{d} \log \Delta N / \mathrm{d} \log \langle N \rangle$ by differentiating a high-order polynomial fitted to the data in Fig.~\ref{fig:GNFs}a. From this data, we operationally define the \emph{apparent} GNF exponent $\aapp$ as the maximum in $\mathrm{d} \log \Delta N / \mathrm{d} \log \langle N \rangle$, located at $R \equiv R^*$, as shown  in Fig.~\ref{fig:GNFs}b. Based on this analysis, in Fig.~\ref{fig:GNFs}c we plot $\aapp$ as a function of $n$, demonstrating the sharp onset of non-Gaussian concentration fluctuations at $n = n_c$. Even given the limitations of its operational definition, it is clear that $\aapp$ does not approach any universal value, as it depends strongly on both $\ell_p$ and on $n$, with a maximum slightly above $n_c$. Unsuprisingly, $\aapp$ also exhibits strong finite size effects, although it asymptotes to a characteristic value for each $\ell_p$ as $H$ becomes sufficiently large~\citep{SI}. 

To investigate the origin of the non-monotonic behaviour in Fig.~\ref{fig:GNFs}a, in Fig.~\ref{fig:GNFs}d we plot $R^*$, which characterises the length scale at which density fluctuations are maximised, as a function of the system size $H$. For true GNFs, one expects the region where non-Gaussian fluctuations are observed to grow in an unbounded fashion with $H$, approaching a universal power-law in the thermodynamic limit. Here, $R^*$ instead exhibits an initial increase until it saturates at $R^* \approx 0.8 \ell_p$ for $H \approx 20 \ell_p$. (The suboptimal data collapse for $\ell_p = 15$ is an inertial effect, which we discuss further below.) This analysis clearly shows that the non-Gaussian fluctuations which emerge above $n_c$ are not ``true'' GNFs in the sense of being a universal property that survives in the thermodynamic limit, but rather ``pseudo-GNFs'' that occur transiently over mesoscopic length scales. These results establish that $R^*$ is directly proportional to $\ell_p$ above the transition to collective motion, although their relationship contains a density-dependent prefactor which we expect to remain close to unity reasonably far above $n_c$. 

\begin{figure}[h!]
\centering
 \subfloat{\includegraphics[width=75mm, clip, trim=0.4cm 0.6cm 0.4cm 0.4cm]{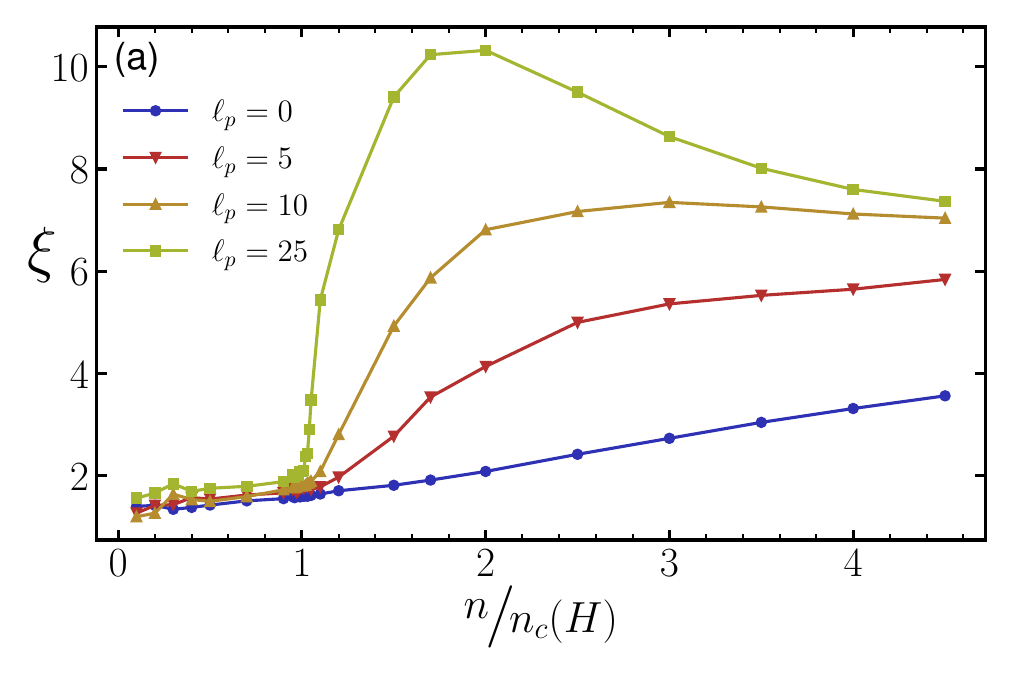}} \\
 \subfloat{\includegraphics[width=75mm, clip, trim=0.4cm 0.4cm 0.4cm 0.4cm]{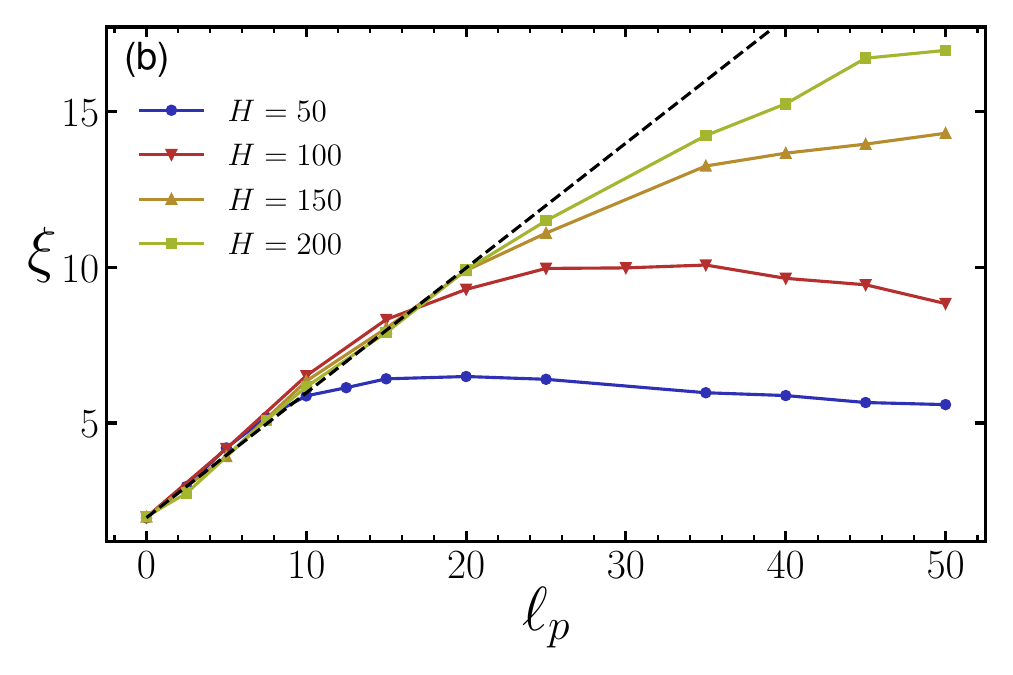}}
 \caption{(a) Fitted values of the nematic length-scale $\xi$ as a function of the reduced microswimmer density for $H = 100$ and various $\ell_p$, as indicated. Notice the similarity with the data in Fig.~\ref{fig:GNFs}c. (b) $\xi$ plotted as a function of $\ell_p$ for $n / n_c = 2.0$ and four different values of $H$, as indicated. The dashed line shows a linear fit to the first six data points for $H = 200$.}\label{fig:xi}
\end{figure}

In dry active systems, GNFs arise due to the coupling between (quasi-)long range  orientational order and non-equilibrium dynamics~\citep{Sriram:EPL:2003,Sriram:AnnuRev:2010}. In the presence of a fluid, the emergent, orientationally ordered state described by the instability criterion of Eq.~\eqref{eq:n_c_inf} is however unstable for finite wavenumbers~\citep{Simha:PRL:2002}, resulting in the chaotic motion that is characteristic of the turbulent state. To establish a connection between local nematic order and the pseudo-GNFs established in Fig.~\ref{fig:GNFs}, we calculate the distance-dependent, scalar swimmer-swimmer nematic order $S(r) \equiv \frac{1}{2}\langle 3 \cos^2 \theta -1 \rangle_r$, where $\theta$ is the angle formed between the two swimmer orientation vectors and $\langle \cdots \rangle_r$ denotes a spatiotemporal average over all swimmer pairs with a given separation $r$. For $\ell_p = 0$, we find that $S(r)$ is well fitted by a stretched exponential, $S(r) = S_0 e^{-(r/\xi)^\beta}$ with a single decay length $\xi$ and a weakly density-dependent $\beta$ ranging between 1.05 and 1.5~\citep{SI}. For $\ell_p > 0$, $S(r)$ has a complex shape with no obvious analytical description~\citep{Bardfalvy:SoftMatter:2019}, and we therefore instead fit the initial decay of $S(r)$ using a regular exponential form, $S(r) = S_0 e^{-r/\xi}$~\citep{SI}. In spite of this complexity, for both shakers and swimmers it is natural to interpret $\xi$ as the characteristic linear size of nematic patches in the turbulent state. In Fig.~\ref{fig:xi}a we show the extracted values of $\xi$ as a function of $n/n_c$ for the same parameter set as in Fig.~\ref{fig:GNFs}c. Even though the jump in $\xi$ at $n = n_c$ is less sharp than for $\aapp$, the similarity between the two datasets is striking, and it is clear that the onset of collective motion for large $\ell_p$ is characterised by a sharp increase in the range of nematic order. In Fig.~\ref{fig:xi}b we further plot $\xi$ as a function of $\ell_p$ for $n / n_c = 2.0$ and four different $H$. Even though finite-size effects are large, the four datasets all present the same linear scaling of $\xi$ on $\ell_p$, although the large-$\ell_p$ behaviour is likely also  affected by swimmer inertia, as discussed below. This result strongly suggests that the size of nematic patches in an unbounded system at a given $n / n_c > 1$ is directly controlled by $\ell_p$. 
This finding might initially seem surprising, since the most unstable wave-number of the homogeneous and isotropic state encoded in Eq.~\eqref{eq:n_c_H} corresponds to the largest available length-scale in the system~\citep{Skultety:JFM:2024}. However, $\xi$ is more likely controlled by the (finite) smallest stable wavenumber of the \textit{nematic} state, which we expect to set the maximum size of stable nematic patches in the turbulent state, similar to what has been found previously for continuum active nematics~\citep{Martinez_Prat:NatPhys:2019}. In End Matter, we show how, starting from a mean-field description of the microswimmer suspension in terms of the relevant mesoscopic order parameter fields ~\citep{ViktorThesis,Baskaran:PNAS:2009,Saintillan2015}, the persistence length $\ell_p$ and tumbling time $\lambda^{-1}$ emerge as the natural length and time scales for nondimensionalising the resulting equations. When rescaled in this manner, these equations depend on \textit{a single} dimensionless control parameter, namely $n / n_c$. By this argument, \textit{any} lengthscale $\tilde{\ell}^*$ emerging as a result of interactions between microswimmers can be written in dimensionless form as $\tilde{\ell}^* = f(n / n_c)$, where $f$ is an unknown function, so that the corresponding dimensional length scale becomes ${\ell}^* = \ell_p f(n / n_c)$. This implies that $\ell_p$ sets the scale for \textit{all} emergent properties of the collective motion state, including the linear size of nematic patches as measured through $\xi$ in Fig.~\ref{fig:xi}b

The results presented in this Letter thus establish a clear connection between (\textit{i}) the characteristic length scale over which non-Gaussian concentration fluctuations are observed, as encoded in $R^*$, and (\textit{ii}) the linear size of nematic patches in the turbulent regime, as encoded in $\xi$. We observe that number fluctuations inside a nematic patch grow until reaching $R^* \approx \xi$, where the latter length scale is selected by the largest stable length scale of a nematically ordered state in the presence of fluid flows~\citep{Sriram:EPL:2003}. At scales much larger than $\xi$, the orientational correlations between individual patches vanish~\citep{Bardfalvy:SoftMatter:2019}, restoring Gaussian fluctuations as in equilibrium. Thus, the non-Gaussian number fluctuations observed in our model do not constitute an example of GNFs in the universal sense, which is a direct consequence of the absence of a stable, nematically ordered state. Since this instability is generic in the presence of fluid flows and occurs for arbitrary values of activity, whether extensile or contractile~\citep{Simha:PRL:2002}, our results suggest that true GNFs are in fact \textit{generically} absent from nematically aligning, momentum-conserving active systems. Most importantly, this conclusion includes wet active nematics~\citep{Doostmohammadi:NatComm:2018}, for which most previous studies of active turbulence have been conducted. The primary difference between this class of continuum models and the microswimmer suspensions studied here is the origin of the initial instability of the isotropic state. In microswimmers, this stems from the flow alignment Eq.~\eqref{eq:pdot}, while for an active nematic it is typically (although not exclusively~\citep{Santhosh:JSTAT:2020}) caused by an underlying free energy that induces an isotropic-nematic transition even in the absence of activity. Since the ensuing collective motion state is fundamentally caused by the same instability~\citep{Simha:PRL:2002} in both models, our conclusions about the absence of true GNFs holds equally for both classes of systems.

Our results show that both $\xi$ and $R^*$ are directly proportional to the microscopic persistence length $\ell_p$, showing that, even though $v_s$ is absent from the instability criterion in Eq.~\eqref{eq:n_c_inf}, swimming has a profound effect on the properties of both the pretransitional suspension~\citep{Skultety:PRX:2020} and the turbulent state. Since $R^* \sim \xi \sim \ell_p$, there exists a possibility of restoring  universal GNFs in the limit $\ell_p \rightarrow \infty$: Because $\aapp$ is strictly bounded by unity the results in Fig.~\ref{fig:GNFs} suggest that, for sufficiently large $\ell_p$ and $H \gg \ell_p$, one might recover an extended power-law region with a universal exponent. Extending the numerics to larger $\ell_p$ is however computationally prohibitive, for two reasons: First of all, reaching $H \gg \ell_p$ requires massive system sizes, which limits the data range in Fig.~\ref{fig:GNFs}d. Secondly, since the LB method intrinsically operates at finite Reynolds number, increasing $v_s$ eventually leads to non-negligible effects of swimmer inertia, which become significant for $\mathrm{Re} \equiv v_s \ell_p \rho / \mu \geq 0.1$~\citep{Nash:PRE:2008,deGraaf:PRE:2017}. For the current parameter set, this translates to $\ell_p \geq 9$, in agreement with the observed decline in the data collapse in Fig.~\ref{fig:GNFs}d. An alternative approach is a field-theoretical treatment of the emergence of non-Gaussian fluctuations, similar to the strategy used in the dry case~\citep{Sriram:EPL:2003,Marchetti:PRE:2018,Toner:JCP:2019}. A direct translation of this approach to wet systems is however not possible even at linear level, as it relies on the long-wavelength perturbation of a stable, nematically ordered base state, which is absent in the presence of fluid flows. 

Regardless of whether or not true GNFs are formally restored in the $\ell_p \rightarrow \infty$ limit, our data clearly establish that GNFs in the universal sense cannot not be generically expected in wet active systems with underlying nematic order. Our results thus provide a new perspective on observations of non-Gaussian fluctuations in biological realisations of active matter~\citep{Cheng:SoftMatter:2021,Silberzan:SoftMatter:2014,Sano:PRE:2017,Peruani:PRL:2012,Swinney:PNAS:2010} as well as novel insights into the factors that control the emergent properties of collective motion in bacterial suspensions. 

\textit{Acknowledgements.} Helpful discussions with Sriram Ramaswamy, Ananyo Maitra, and Eric Cl\'ement are kindly acknowledged. JS acknowledges funding from the Knut and Alice Wallenberg Foundation (grant 2019.0079) and the Swedish Research Council (grant 2019-03718). 
\bibliography{bibliography}

\begin{thebibliography}{42}
\bibitem [{SI()}]{SI}%
  \BibitemOpen
  \href@noop {} {\bibinfo {title} {See {S}upplemental {M}aterial at [{URL} to
  be inserted by publisher] for further details.}}\BibitemShut {Stop}%
\end{thebibliography}

\appendix

\section*{End Matter}
\noindent \textit{Simulation Details.} In terms of LB units, defined by the lattice spacing $\Delta L$ and the time step $\Delta t$, the swimmer parameters were set to $F = 1.57 \times 10^{-3}$, $\ell = 1$, roughly corresponding to the value of $\kappa$ measured for \emph{E.~coli}~\citep{Goldstein:PNAS:2011,Bardfalvy:SoftMatter:2019}, and $\lambda = 2 \times 10^{-4}$. We furthermore set $\mu = 1/6$, corresponding to full relaxation of the fluid to local equilibrium at each timestep, and $\rho = 1$. As we showed in~\citep{Skultety:PRX:2020}, a bulk microswimmer suspension is described by two control parameters: the reduced microswimmer density, $n / n_c^{\infty}$, where $n_c^{\infty}$ is given by Eq.~\eqref{eq:n_c_inf}, and the swimming persistence length $\ell_p = v_s / \lambda$. In order to keep $n_c^{\infty}$ constant, $\ell_p$ was controlled by varying the swimming speed $v_s$ between zero and $5 \times 10^{-3}$ while keeping $\lambda$ constant. All simulations were initialised from a configuration with random swimmer positions and orientations and all the data acquired before reaching steady state, as monitored through the time dependence of the fluid velocity variance $\langle U^2 \rangle$, was discarded from the analysis. Unless otherwise stated, results are presented for a system size $H = 100$ and all lengths are expressed in terms of the swimmer length $\ell$. \\

\begin{figure}[h]
  \begin{center}
  \includegraphics[trim=0.1cm 1.0cm 1.93cm 0.45cm, clip, width=60mm]{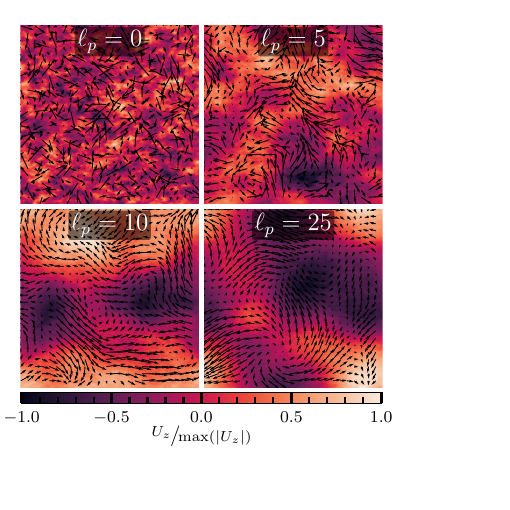}
\caption{Snapshots of a $2d$ slice of the $3d$ flow field for $n / n_c = 2.0$ and $H = 250$, for the values of $\ell_p$ indicated. Arrows show the in-plane component and colours show the out-of-plane component of $\UU$, scaled by its maximum value.}\label{fig:snapshots}
  \end{center}
\end{figure}

\noindent \textit{$\ell_p$-dependence of collective flow fields.} Figure~\ref{fig:snapshots} shows simulation snapshots of the flow field $\UU$ deep in the collective motion state. These clearly show the strong effect of changing $\ell_p$ on the appearance and spatial correlations of the collective flow fields, which is also reflected in the increased range of nematic swimmer-swimmer correlations, as discussed in the main text.

\noindent \textit{Mean-field description of the suspension dynamics.} Our starting point is a mean-field kinetic theory similar to those developed previously in the literature to describe dilute suspensions of dipolar, needle-like  microswimmers~\citep{Baskaran:PNAS:2009,Saintillan2015,Saintillan:PoF:2008}. While differing in technical details, all these theories encode the following simple microscopic dynamics: each particle swims in the direction of its orientation with the swimming speed $v_s$, while randomly changing its orientation (tumbling) with a rate $\lambda$. Additionally, each microswimmer generates a long-ranged dipolar velocity field that advects and re-orients all other microswimmers. The resulting kinetic theory allows one to identify the microswimmer density $c(\bm r,t)$, their average polarisation $\bm m (\bm r, t)$, and the nematic order parameter $\bm Q(\bm r, t)$ as the crucial degrees of freedom to describe the transition to collective motion in such systems. Employing the same methods previously utilised in this context~\citep{Saintillan2015,Baskaran:PNAS:2009} we expand the one-particle distribution function of the kinetic theory in terms of its orientational moments (see~\citep{ViktorThesis} for details). Enslaving the third order moment to the dynamics of $c$, $\bm m$, and $\bm Q$ and retaining only the leading order in the fields and their derivatives, we obtain the following equations of motion for $c$, $\bm m$, and $\bm Q$ in 3 dimensions:

	\begin{align}
		\left(\partial_t + U_\alpha \nabla_\alpha \right) c =& - \nabla_{\alpha} m_{\alpha}, \\
		\left(\partial_{t} + U_\alpha \nabla_\alpha \right) m_{i} =& - m_{i} - \nabla_{\alpha} (Q_{i \alpha} + \frac{1}{3} \delta_{i \alpha} c) \nonumber \\
		&\ + \frac{4}{5} m_{\alpha} \nabla_{\alpha} U_{i} - \frac{1}{5} m_{\alpha} \nabla_{i} U_{\alpha}, \\
		\left(\partial_{t} + U_\alpha \nabla_\alpha \right) Q_{ij} =&  - Q_{ij} \nonumber \\
		&\ - \frac{1}{5} (\nabla_{j} m_{i} + \nabla_{i} m_{j} - \frac{2}{3} \delta_{ij} \nabla_{\alpha}m_{\alpha} ) \nonumber \\
		&\ + \frac{1}{5} c (\nabla_{j}U_{i} + \nabla_{i}U_{j}) + \frac{1}{7}\nabla^2 Q_{ij} \nonumber \\
        &\ + \frac{3}{35}\Big( \nabla_i\nabla_\alpha Q_{\alpha j} + \nabla_j\nabla_\alpha Q_{\alpha i} \nonumber \\
        &\ - \frac{2}{3}\delta_{ij} \nabla_\alpha\nabla_\beta Q_{\alpha \beta}\Big) \nonumber \\
        &\ - \frac{2}{7} \delta_{ij} Q_{\alpha\beta} \nabla_{\alpha} U_{\beta} \nonumber \\
		&\ + \frac{5}{7} \left(Q_{i\alpha}\nabla_{\alpha}U_{j}+ Q_{j\alpha}\nabla_{\alpha}U_{i}\right) \nonumber \\
		&\ - \frac{2}{7} (Q_{i\alpha}\nabla_{j}U_{\alpha} + Q_{j\alpha}\nabla_{i}U_{\alpha}), \\
	        \nabla_{i} (p + \frac{1}{3}c) = &\  \nabla^{2} U_{i} - 5\frac{n}{n_c} \nabla_{\beta} Q_{i\beta}, \label{eq:Stokes}\\
		 \nabla_{i} U_{i} = &\ 0.
	\end{align}
Here, $p$ and $\bm U$ are, respectively, the fluid pressure and velocity, indices denote Cartesian spatial coordinates, and we assume summation over repeated indices. The crucial point is that these equations have been made dimensionless based on the following rescalings: (\textit{i}) time is rescaled by $\lambda^{-1}$, (\textit{ii}) length is rescaled by $\ell_p = v_s / \lambda$, (\textit{iii}) fluid velocity is rescaled by $v_s$, and (\textit{iv}) $c$, $\bm m$, and $\bm Q$ are rescaled by the average microswimmer density, $n$. As can be seen by visual inspection, when using this nondimensionalisation, the suspension's dynamics are controlled by \textit{a single} dimensionless parameter occuring in the Stokes equation (Eq.~\eqref{eq:Stokes}), namely $n / n_c$, which encodes the distance from the onset of active turbulence, with $n_c$ for an infinite system given by Eq.~\eqref{eq:n_c_inf}. As discussed in the main text, a crucial consequence of this observation is that \textit{any} lengthscale $\tilde{\ell}^*$ emerging as a result of interactions between microswimmers can be written in dimensionless form as a function of this single parameter, \textit{i.e.}, $\tilde{\ell}^* = f(n/n_c)$, where $f$ is an unknown function.

\end{document}